\documentstyle[amsfonts,12pt]{article}
\begin{document}
\newcommand{\pr}{{\rm pr}}
\newcommand{\vep}{\varepsilon}
\renewcommand{\theequation}{\thetheorem}
\newtheorem{theorem}{Theorem}[section]
\newtheorem{corollary}[theorem]{Corollary}
\newtheorem{lemma}[theorem]{Lemma}
\newtheorem{definition}[theorem]{Definition}
\newtheorem{conjecture}[theorem]{Conjecture}
\newtheorem{proposition}[theorem]{Proposition}
\title{Antichaos in a Class of Random Boolean Cellular Automata}
\author{James F. Lynch
\thanks{Research supported by NSF Grant
CCR-9006303.} \\
Department of Mathematics and Computer Science \\
Clarkson University \\
Potsdam, N. Y. 13699-5815}
\date{}
\maketitle

\begin{center} Abstract \end{center}

{\em A variant of Kauffman's model of cellular metabolism is presented.
It is
a randomly generated network of boolean gates, identical to Kauffman's
except for a small bias in favor of boolean gates that depend on at most
one input. The bias is asymptotic to 0 as the number of gates increases.
Upper bounds on the
time until the network reaches a state cycle and the size of
the state cycle, as functions of the number of gates $n$, are derived.
If the bias approaches 0 slowly enough, the state cycles will be
smaller than $n^c$ for some $c<1$. This lends support to Kauffman's claim
that in his
version of random network the average size of the state cycles is
approximately $n^{1/2}$.} \hfill \\
\hspace*{\fill} \\
Proposed running head: {\bf Antichaos in Random Boolean Nets} \hfill \\
\hspace*{\fill} 
\section{Introduction}

Let $n$ be a natural number. A {\em boolean cellular automaton}
with $n$ gates consists
of a directed graph $D$ with vertices $1,\ldots,n$ (referred to as {\em
gates})
and a sequence $b = (b_1,\ldots,b_n)$ of boolean functions.
The number of arguments of each function $b_i$ is the same as the indegree
of gate $i$. We say that gate $j$ is an {\em input} to gate $i$ if
$(j,i)$ is an edge of $D$. A boolean cellular automaton $B=
\langle D,b\rangle$ defines a map from $\{0,1\}^n$ (the set of 0-1 sequences of
length $n$) to $\{0,1\}^n$ in the following way. For each $i=1,\ldots,n$ let
$j_{i1},\ldots,j_{ik_i}$ be the inputs of $i$ in increasing order.  Given $x =
(x_1,\ldots,x_n) \in \{0,1\}^n$, $B(x) =
(b_1(x_{j_{11}},\ldots,x_{j_{1k_1}}),\ldots,b_n(x_{j_{n1}},\ldots,
x_{j_{nk_n}}))$.
$B$ may be regarded as a finite state automaton with state set $\{0,1\}^n$
and initial state $x$. That is, its state at time 0 is $x$, and if its
state at time $t$ is $y \in \{0,1\}^n$ then its state at time $t+1$ is
$B(y)$. We put $B^t(x)$ for the state of $B$ at time $t$, and
$b_i^t(x)$ for the value of its $i$th component, or gate, at time $t$.
Since the number of states is finite, i.e. $2^n$, there exist times
$t_0$ and $t_1$ such that $B^{t_0}(x) = B^{t_1}(x)$. Let $t_1$ be the
first time at which this occurs. Then $B^{t+t_1-t_0}(x) = B^t$ for all
$t \geq t_0$. We refer to the set of states $\{B^t(x) : t \geq t_0 \}$
as the {\em state cycle\/} of $x$ in $\langle D,b \rangle$, to
distinguish it from a cycle of $D$ in the graph-theoretic sense. The
{\em tail\/} of $x$ in $\langle D,b \rangle$ is $\{B^t(x) : t < t_0
\}$.

S. Kauffman \cite{k0} proposed boolean cellular automata as models of
cellular metabolism. The gates represent genes within a cell, the state
of a gate indicates whether the gene is active or inactive, and
$\langle D,b \rangle$ describes how the activity of genes affects other
genes. State cycles correspond to the possible behavior patterns that
the cell can differentiate into.

In an extensive series of articles (see for example \cite{k0,k1,k2}),
Kauffman described computer simulations on a particular kind of randomly
constructed boolean cellular automaton. The number of gates $n$ was fixed,
typically in the range of several hundred to several thousand. Every gate
had exactly two inputs, chosen independently with equal probability from
the $\left(\begin{array}{c} n \\2 \end{array}\right)$ possibilities.
Each gate was also assigned a random boolean function on its two inputs from
the 16 equally likely possibilities. Then a random starting state was chosen,
again with a uniform distribution on $\{0,1\}^n$. The automaton was
simulated, and the trajectory of its state at discrete time intervals was
observed.

A striking feature of the behavior of these random boolean cellular
automata was their stability. Typically, the tail length
and the size of the state cycle were quite small compared to $n$. Kauffman
estimated the
median size of the state cycle was on the order of $n^{1/2}$. Also, the
automata were very robust. Perturbing the state by flipping the value of
one gate usually did not affect the state cycle that was entered.

There are a number of interesting biological implications of these
simulations. Perhaps the most fundamental is that there may be another
major influence on the evolution of species besides natural selection.
Small, stable state cycles are necessary to the proper functioning of
any cell. Mutations are random modifications to the genes that control
the cell. Kauffman's simulations suggest that most mutations preserve
orderly behavior, and consequently may be passed on. If almost all
mutations were selected against, evolution would be extremely slow. The
evidence that many randomly constructed boolean cellular automata
exhibit antichaotic behavior supports the idea of modern evolutionary
theory that chance plays a greater role and natural selection a lesser
role in evolution than previously thought.

Although Kauffman's model has been studied empirically for many years,
it was only recently that formal mathematical methods have been applied
to it (see for example \cite{f,lc}). {\L}uczak and Cohen appear to be
the first to prove one of Kauffman's claims--that perturbation of a
single gate almost never affects the state cycle that the automaton enters.
They also derived a $2^{o(n)}$ upper bound on the size of the state cycle.

In this article, we study a random boolean cellular automaton that is
almost the same as Kauffman's (we shall make this precise shortly).
Essentially, when boolean functions are assigned to the gates,
we give a very small bias to those functions that depend on at most
one argument. Depending on the strength of the bias,
we will derive polynomial upper bounds on the
size of the state cycle entered, and even sublinear upper bounds. We also
obtain sublinear bounds of the tail length.
\section{Preliminaries}
We will use the following notions, introduced by Kauffman \cite{k2}.
\begin{definition} Let $f:\{0,1\}^k \rightarrow \{0,1\}$ be a boolean
function of $k$ arguments. Then $f$ is said to be {\em canalyzing\/} if
there is some $m = 1,\ldots,k$ and some values $u,v \in \{0,1\}$ such that
for all $x \in \{0,1\}^k$, if $x_m = u$ then $f(x) = v$. Argument $m$ of $f$
is said to be a {\em forcing argument\/} with {\em forcing value\/} $u$ and
{\em forced value\/} $v$. Likewise, if $\langle D,b\rangle$ is a boolean
cellular automaton and $b_i$ is a canalyzing function with forcing argument
$m$, forcing value $u$ and forced value $v$, then input $j_{im}$ is a {\em
forcing input\/} of gate $i$.  That is, if the value of $j_{im}$ is $u$ at time
$t$, then the value of $i$ is guaranteed to be $v$ at time $t+1$.
\end{definition}
Note that all four 1-input boolean functions are trivially canalyzing, and
all of the 2-input boolean functions except equivalence and exclusive or
are canalyzing.

The next definition is borrowed partly from Kauffman \cite{k2} and
{\L}uczak and Cohen \cite{lc}.
\begin{definition} Again, $\langle D,b\rangle$ is a boolean cellular automaton.
Using induction on $t$, we define what it means for gate $i$ to be {\em forced
to a value\/} $v$ {\em in\/} $t$ {\em steps}.

If $b_i$ is the constant function $v$, then $i$ is forced to $v$ in 0 steps.

If all inputs $j_{i1},\ldots,j_{ik}$ of $i$ are forced to $u_1,\ldots,u_k$
respectively in $t$ steps then $i$ is forced to $b_i(u_1,\ldots,u_k)$
in $t+1$ steps.

If $b_i$ is a canalyzing function with forced input $m$, forcing value $u$,
and forced value $v$, and $j_{im}$ is forced to $u$ in $t$ steps, then
$i$ is forced to $v$ in $t+1$ steps.
\end{definition}
By induction on $t$ it can be seen that if $i$ is forced in $t$ steps, then
it stabilizes for all initial states $x$ in $t$ steps. That is, for all
$t^{\prime} \geq t$, $b_i^{t^{\prime}}(x) = b_i^t(x)$. However, the converse
is not true. It is easy to construct boolean cellular automata without
any forced gates but with gates
that stabilize for all initial states
(see Figure
1).
\vfill
\noindent
An example of a boolean cellular automaton without any forced gates, all
of whose gates become stable. The labels in the circles denote the
boolean functions assigned to the gates: $\oplus$ is
exclusive or, and $\neg$ is
negation.
\begin{center}
FIGURE 1
\end{center}
\newpage

The next definition is due to {\L}uczak and Cohen \cite{lc}.
\begin{definition}
For any gate $i$ in $\langle D,b\rangle$, let
\begin{eqnarray*}
N_0^-(i) & = & \{i\} \mbox{ and} \\
N_{d+1}^-(i) & = & \bigcup\{N_d^-(j) : j \mbox{ is an input to } i \}.
\end{eqnarray*}
Then
$$
S_d^-(i) = \bigcup_{c \leq d} N_c^-(i).
$$
\end{definition}
Note that whether $i$ is forced in $d$ steps is completely determined by
the
restriction of $D$ and $b$ to $S_d^-(i)$.

The class of {\em random} boolean cellular automata studied in this paper is
the same as Kauffman's except that the probabilities of assigning boolean
functions to gates are slightly biased in favor of the six functions that
depend on at most one argument:
\begin{eqnarray*}
f(x_1,x_2) & = & x_1, \\
f(x_1,x_2) & = & \neg x_1, \\
f(x_1,x_2) & = & x_2, \\
f(x_1,x_2) & = & \neg x_2, \\
f(x_1,x_2) & = & 0\mbox{, and} \\
f(x_1,x_2) & = & 1. \\
\end{eqnarray*}
Let $\vep(n)$ be a function on the natural numbers such that $0 \leq \vep(n)$
for all $n$ and $\lim_{n \rightarrow \infty}\vep(n) = 0$. (We will impose
further conditions on $\vep$ later.) A directed graph $D$ with $n$ gates is
generated as in Kauffman's model, and the sequence $b$ of boolean
functions is generated using the distribution
$$
\mbox{probability that }b_i = f \mbox{ is }
\left\{\begin{array}{ll} 1/16 - \vep(n)/16 & \mbox{ if $f$ depends on both
arguments} \\
1/16 + \vep(n)/16 & \mbox{ if $f$ depends on one argument} \\
1/16 + 3\vep(n)/16 & \mbox{ if $f$ is a constant
function.}\end{array}\right.
$$
The particular coefficients of $\vep(n)$ in the above definition were
chosen to simplify the notation in our proofs. Our theorems apply to any
class of random boolean cellular automata where the probability of
assigning $f$ to gate $i$  is $1/16 - a_f\vep(n)$ for some positive
constant $a_f$ when $f$ depends on both arguments, and the probability
is $1/16 + a_f\vep(n)$ otherwise. The proofs for the general case are
more cumbersome, but they do not involve any important additional
ideas.
We will use $\tilde{B} = \langle
\tilde{D},\tilde{b}\rangle$ to denote a random boolean cellular
automaton generated as
above. For any properties $\cal P$ and $\cal Q$ pertaining to boolean cellular
automata, we put $\pr({\cal P},n)$ for the probability that the random boolean
cellular automaton on $n$ gates has property $\cal P$ and $\pr({\cal P}|
{\cal Q},n)$ for the conditional probability that $\cal P$ holds, given that
$\cal Q$ holds. Usually, we will omit the $n$ in these expressions since it
will be understood.

The properties of $\langle D,b \rangle$ that we are really interested in
depend only on the function of $\{0,1\}^n$ computed by $\langle D,b
\rangle$. What this means is that if $j_1$ and $j_2$ are the two inputs
of a gate $i$ and $b_i = f$, where $f$ depends on only one argument, say
$f(x_1,x_2) = x_1$, then we can delete the edge $(j_2,i)$ from $D$ and
replace $f$ by the one argument function $g(x_1) = x_1$. In this case we
say that the input of $i$ is $j_1$ and $b_i = g$. If $f$ is a constant
function, we could delete both input edges, but for our purposes it is
simpler just to delete the second input edge and regard $f$ as a
constant function of $x_1$. This will make our proofs easier, and is the
motivation for defining a slightly different class of random boolean
cellular automata, which is essentially equivalent to the class just
defined.

We take $\vep$ and $n$ as before. For $i = 1,\ldots,n$
independently, gate $i$ will have two inputs with probability $1-\vep(n)$ and
it will have one input with probability $\vep(n)$. If it has two inputs then
the inputs and the boolean function assigned to it are chosen as in Kauffman's
model. If it has one input then the input is chosen with uniform probability
from among the $n$ gates, and its boolean function is chosen from the four
possibilities with equal probability. We will use $\tilde{B}^{\prime} = \langle
\tilde{D}^{\prime},\tilde{b}^{\prime}\rangle$ to denote a random boolean
cellular
automaton generated this way, and $\pr^{\prime}$ for its associated
probability function. The following proposition shows that these two
models of random boolean cellular automata have the same probability
distribution with respect to the functions that they
compute.
\begin{proposition}
For any natural number $n$, let $\tilde{B} = \langle \tilde{D},\tilde{b}
\rangle$
and $\tilde{B}^{\prime} = \langle \tilde{D}^{\prime},\tilde{b}^{\prime}
\rangle$
be the two random boolean cellular automata with $n$ gates defined above.
Then for any function $F : \{0,1\}^n \rightarrow \{0,1\}^n$,
$\pr(\tilde{B} = F) = \pr^{\prime}(\tilde{B}^{\prime} = F)$.
\end{proposition}
{\bf Proof.} We need only show the following.
\begin{enumerate}
\item For every gate $i$, every $1 \leq j_1<j_2 \leq n$, and every
boolean function $f$ that depends on both arguments,
$$
\pr^{\prime}(\mbox{inputs of $i$ are $j_1,j_2$ and
$b_i^{\prime} = f$}) =
\pr(\mbox{inputs of $i$ are $j_1,j_2$ and $b_i = f$}).
$$
\item For every gate $i$, every $1 \leq j \leq n$, and every boolean
function $g$ of one argument,
$$
\pr^{\prime}(\mbox{input of $i$ is $j$ and
$b_i^{\prime} = g$}) =
\pr(\mbox{input of $i$ is $j$ and $b_i = g$}).
$$
\item For every gate $i$ and constant boolean
function $g$,
$$
\pr^{\prime}(b_i^{\prime} = g) =
\pr(b_i = g).
$$
\end{enumerate}
To prove 1.,
\begin{eqnarray*}
\pr^{\prime}(\mbox{inputs of $i$ are $j_1,j_2$ and
$b_i^{\prime} = f$}) & = & (1 - \vep(n)) \times \frac1
  {\left(\begin{array}{c} n \\ 2 \end{array}\right)} \times \frac1{16}
  \\
  & = & \frac1{\left(\begin{array}{c} n \\ 2 \end{array}\right)}
  \times \left(\frac1{16} - \frac{\vep(n)}{16} \right) \\
 & = & \pr(\mbox{inputs of $i$ are $j_1,j_2$ and $b_i = f$}).
\end{eqnarray*}
To prove 2.,
\begin{eqnarray*}
\pr^{\prime}(\mbox{input of $i$ is $j$ and
$b_i^{\prime} = g$}) & = & (1 - \vep(n)) \times \frac{n-1}
  {\left(\begin{array}{c} n \\ 2 \end{array}\right)} \times \frac1{16}
  + \vep(n) \times \frac1n \times \frac14 \\
  & = & \frac2n \times \left( \frac1{16} + \frac{\vep(n)}{16} \right)
  \\
 & = & \pr(\mbox{input of $i$ is $j$ and $b_i = g$}).
\end{eqnarray*}
To prove 3.,
\begin{eqnarray*}
\pr^{\prime}(b_i^{\prime} = g) & = & (1 - \vep(n)) \times \frac1{16}
  + \vep(n) \times \frac14 \\
  & = & \frac1{16} + \frac{3\vep(n)}{16} \\
 & = & \pr(b_i = g).
\end{eqnarray*}
\hspace*{\fill} $\Box$ \\
Thus we make no distinction between
$\langle \tilde{D},\tilde{b} \rangle$ and
$\langle \tilde{D}^{\prime},\tilde{b}^{\prime} \rangle$.

The following lemma is essentially a generalization of Lemma 1 in \cite{lc}.
\begin{lemma}\label{l1}
For sufficiently large $n$, any natural number $d$, and any gate $i$
in the random boolean cellular automaton $\langle\tilde{D},\tilde{b}\rangle$
with $n$ gates,

$$
\pr(i \mbox{ is not forced in $d$ steps }|\mbox{ $S_d^-(i)$ induces an acyclic
subgraph of $\tilde{D}$}) \leq \frac{16}{d}.
$$
\end{lemma}
{\bf Proof.} Let $p_d$ be the conditional probability in question. We first
show that it satisfies the following recurrence.
\begin{eqnarray*}
p_0 & = & 7/8 - 3\vep(n)/8, \\
p_{d+1} & = & (1-\vep(n)/2)p_d - (1-\vep(n))p_d^2/8.
\end{eqnarray*}
The base case for $p_0$ follows from the fact that two out of the 16 2-input
boolean functions are constant while two out of the four of the 1-input
boolean functions are constant. Thus $p_0 = (1-\vep(n))(7/8) + \vep(n)(1/2)$.

To prove the induction step, first take the case when $i$ has two inputs.
Then $i$ is not forced in $d+1$ steps if and only if neither of its inputs
is forced in $d$ steps and $b_i$ is not constant, or exactly one of its inputs
is forced but its forced value is not forcing for $i$. The first possibility
has probability $p_d^2 \times 7/8$, while the second has probability
$2p_d(1-p_d) \times 1/2$. If $i$ has one input, then it is not forced if and
only if the input is not forced and $b_i$ is not constant. This probability
is $p_d \times(1/2)$. Altogether we have
\begin{eqnarray*}
p_{d+1} & = & (1-\vep(n))(7p_d^2/8 + p_d(1-p_d)) + \vep(n)p_d/2 \\
 & = & (1 - \vep(n)/2)p_d - (1 - \vep(n))p_d^2/8.
\end{eqnarray*}

Let $q_d = 1/p_d$. We will show by induction on $d$ that $q_d \geq d/16$
for sufficiently large $n$,
from which the Lemma follows. Clearly $q_0 \geq 0$.

Assuming $q_d \geq d/16$, we use our recurrence for $p_{d+1}$, getting
$$
1/q_{d+1} = (1 - \vep(n)/2)/q_d - (1 - \vep(n))/(8q_d^2).
$$
Rearranging,
\begin{eqnarray*}
q_{d+1} & = & \frac{q_d}{(1 - \vep(n)/2) - (1 - \vep(n))/(8q_d)} \\
 & \geq & \frac{q_d}{1 - (1 - \vep(n))/(8q_d)} \\
 & \geq & q_d + (1 - \vep(n))/8 \\
 & \geq & q_d + 1/16 \mbox{ for sufficiently large $n$ since $\vep(n)
  \rightarrow 0$.}
\end{eqnarray*}
Thus $q_{d+1} \geq (d+1)/16$, and the proof is complete. \hfill $\Box$

Our final basic idea concerns chains of gates that are not likely to
stabilize. An {\em unforced path\/} is a sequence of distinct gates
$P = (i_1,\ldots,i_p)$ such that $i_r$ is an input of $i_{r+1}$ for
$1 \leq r < p$ and none of the gates are forced in
$256/\vep(n)$ steps. An {\em unforced
cycle\/} is the same except $i_1 = i_p$.
\section{Theorems and Proofs}
In this section, we will prove our upper bounds on the sizes of the
tails and state
cycles of the random boolean cellular automaton
$\langle \tilde{D},\tilde{b}\rangle$ as a function of $\vep$.
First, we prove some upper bounds on the sizes of unforced structures
in $\langle \tilde{D},\tilde{b}\rangle$. Throughout the paper, $\log$
will mean $\log_2$.
\begin{lemma}\label{l2}
If $\vep(n) \gg 1/\log n$ then
$$
\pr(\langle \tilde{D},\tilde{b}\rangle\mbox{\rm \ has an unforced path
longer than }4\log n/\vep(n))
 = o(1).
$$
\end{lemma}
{\bf Proof.} Let $l = \lceil 4\log n/\vep(n)\rceil$. The gates in a path of
length $l$
can be chosen in $n(n-1)\ldots(n-l+1) \leq n^l$ ways. Having chosen the $l$
gates $i_1,\ldots,i_l$, for $r = 1,\ldots,l$ let ${\cal P}_r$ be the
event that $i_1,\ldots,i_r$ form an unforced path. Then
the probability that $i_1,\ldots,i_l$ actually form an unforced
path is bounded above by
$$
\prod_{r=2}^l ((1 - \vep(n))\alpha_r + \vep(n)\beta_r),
$$
where
\begin{eqnarray*}
\alpha_r & = & \pr(\mbox{$i_{r-1}$ is an input to $i_r$ and $i_r$ is not
  forced in $256/\vep(n)$ steps } | \\
  & & \mbox{ indegree$(i_r) = 2$
  and ${\cal P}_{r-1}$}),\\
\beta_r & = & \pr(\mbox{$i_{r-1}$ is the input to $i_r$ and $i_r$ is not
  forced in $256/\vep(n)$ steps } | \\
  & & \mbox{ indegree$(i_r) = 1$ and
   ${\cal P}_{r-1}$}).
\end{eqnarray*}
Clearly $\beta_r = 1/(2n)$.

Assuming $i_r$ has two inputs and ${\cal P}_{r-1}$ holds, let $j_r \neq
i_{r-1}$ be the other input of $i_r$.
Let ${\cal Q}_r$ be the event that
$S_{256/\vep(n)-1}^-(j_r)$ is a tree and $S_{256/\vep(n)-1}^-(j_r) \cap \bigcup
  \{S_{256/\vep(n)}^-(i_s) : s < r\} = \emptyset$.
Then
$$
\alpha_r \leq \left((n-1)/\left(\begin{array}{c} n \\ 2\end{array}\right)
  \right) \times (\gamma_r + \delta_r + \zeta_r)
$$
where
\begin{eqnarray*}
\gamma_r & = & \pr(\mbox{not } {\cal Q}_r), \\
\delta_r & = & \pr(j_r \mbox{ not forced in $256/\vep(n) - 1$ steps }
  |\mbox{ ${\cal Q}_r$}), \\
\zeta_r & = & \pr(i_r \mbox{ not forced  in $256/\vep(n)$ steps }
  |\mbox{ ${\cal Q}_r$ and $j_r$ is forced in $256/\vep(n)-1$ steps}).
\end{eqnarray*}
We now get upper bounds on $\gamma_r$, $\delta_r$, and $\zeta_r$.

First, $S_{256/\vep(n)-1}^-(j_r)$ is not a tree only if there exist two
paths of length at most
$256/\vep(n)-1$ beginning at some common gate and ending at $j_r$.
This probability is bounded above by
\begin{eqnarray*}
\sum_{p=0}^{256/\vep(n)-1} n^p(2/n)^pp^2\sum_{q=1}^{256/\vep(n)-1}
n^{q-1}(2/n)^q & \leq & (256/\vep(n))^42^{512/\vep(n)}n^{-1} \\
 & = & o(\vep(n)) \mbox{ since } 1/\vep(n) \ll \log n.
\end{eqnarray*}
The probability that $S_{256/\vep(n)-1}^-(j_r)\cap\bigcup\{S_{256/
\vep(n)}^-(i_s) : s < r\} \neq
\emptyset$ is bounded above by
$$
l \times 2^{256/\vep(n)} \times \sum_{p=0}^{256/\vep(n)-1} n^{p-1}(2/n)^p
= o(\vep(n)),
$$
so
$$
\gamma_r = o(\vep(n)).
$$

Assuming ${\cal Q}_r$ holds, the event that $j_r$ is not forced in
$256/\vep(n) - 1$ steps is independent of the events that $i_s$ is
forced in $256/\vep(n)$ steps, $s < r$. Therefore by Lemma \ref{l1},
$$
\delta_r \leq 16\vep(n)/255.
$$

Finally, $\zeta_r = 1/2$, so $\alpha_r \leq n^{-1}(1 + 32\vep(n)/255 +
o(\vep(n)))$.
Therefore the probability that $i_1,\ldots,i_l$ form an unforced path is at
most
$$
\{n^{-1}[(1-\vep(n))(1+32\vep(n)/255+o(\vep(n)))+\vep(n)/2]\}^{l-1} \leq
  n^{-l+1}(1 - \vep(n)/4)^{l-1},
$$
and the probability that there exists such $i_1,\ldots,i_l$ is at most
$n(1 - \vep(n)/4)^{l-1}$. Since $l = \lceil 4\log n/\vep(n)\rceil$, this is
asymptotic
to
$$
ne^{-\log n} \rightarrow 0.
$$
\hspace*{\fill} $\Box$
\begin{lemma}\label{l3}
If $\vep(n) \gg 1/\log n$ then
$$
\pr(\langle \tilde{D},\tilde{b}\rangle\mbox{\rm \ has an unforced cycle
larger than } 8\log\log n/\vep(n))
  = o(1).
$$
\end{lemma}
{\bf Proof.} By Lemma \ref{l2}, we need consider only cycles of length
at most $4\log n/\vep(n)$.
Summing over all cycle sizes from $8\log\log n/\vep(n)$ to
$4\log n/\vep(n)$, and using the same estimates as in the proof of Lemma
\ref{l2},
the probability is bounded above by
\begin{eqnarray*}
4\log n/\vep(n) \times (1 - \vep(n)/4)^{8\log\log n/\vep(n)}
 & \leq & 4(\log n)^2 \times e^{-2\log\log n} \\
 & \rightarrow & 0.
\end{eqnarray*}
\hspace*{\fill} $\Box$
\begin{lemma}\label{l4}
If $\vep(n) \gg 1/\log n$ then
$$
\pr(\langle \tilde{D},\tilde{b}\rangle\mbox{\rm \ has unforced cycles
connected by an unforced path})
 = o(1).
$$
\end{lemma}
{\bf Proof.} By Lemmas \ref{l2} and \ref{l3}, we need consider only cycles
of size at most $8\log\log n/\vep(n)$ and paths of length at most
$4\log n/\vep(n)$. The endpoints of the path can be chosen in at most
$(8\log\log n/\vep(n))^2$ ways. Summing over all cycle sizes up to
$8\log\log n/
\vep(n)$, all possible choices of endpoints of the path, and all paths of
length up to $4\log n/\vep(n)$, the probability is bounded above by
$$
(8\log\log n/\vep(n))^4 \times (4\log n/\vep(n))\times n^{-1}
  \rightarrow 0.
$$
\hspace*{\fill} $\Box$
\begin{theorem}
If $\vep(n) \gg 1/\log n$ then
$$
\pr(\langle \tilde{D},\tilde{b}\rangle\mbox{\rm \ has a tail longer than }
9\log n/\vep(n)) = o(1)
$$
\end{theorem}
{\bf Proof.} After $256/\vep(n)$ steps, the only gates that are not yet
stable are those in unforced paths and cycles. We may assume that all
the cycles and paths are disjoint except possibly at the endpoints of
the paths. By Lemma \ref{l4}, with probability $1-o(1)$, no path begins
and ends at a cycle (see Figure 1.(a)).

Let $l$ be the length of the longest unforced path in $\langle
\tilde{D},\tilde{b} \rangle$ and $m$ be the size of its largest unforced
cycle. After $l$ more steps, the only gates that are not yet stable are
those in unforced cycles and paths beginning at an unforced cycle (see
Figure 1.(b)).

Now consider the state of the gates in these cycles, i.e. the projection
of the state of $\langle \tilde{D},\tilde{b} \rangle$, where we look
only at the values of the gates in the unforced cycles. This state will
reach its state cycle in at most $2m$ steps. Then $\langle
\tilde{D},\tilde{b} \rangle$ will reach its state cycle in at most $l$
more steps. The Theorem then follows from Lemmas \ref{l2} and \ref{l3}.
\hfill $\Box$
\noindent
Typical structure of the unstable gates of $\langle
\tilde{D},\tilde{b} \rangle$. \hfill \\
(a) After $256/\vep(n)$ steps. \hfill \\
(b) After $l$ more steps, where $l$ is the maximum length of unforced
paths that do not begin at an unforced cycle.
\begin{center}
FIGURE 2
\end{center}
\begin{corollary}
If $\vep(n) \gg 1/\log n$ then
$$
\lim_{n \rightarrow \infty} \pr(\mbox{\rm all tail lengths of }\langle
\tilde{D},\tilde{b} \rangle \mbox{\rm \  are } o((\log n)^2)) = 1.
$$
\end{corollary}
\begin{theorem}
If $\vep(n) \gg 1/\log n$ then
$$
\pr(\langle \tilde{D},\tilde{b}\rangle \mbox{\rm \ has a state cycle larger
than }
2^{12\log\log n/\vep(n)}) = o(1).
$$
\end{theorem}
{\bf Proof.} Any path (or cycle) of unstable gates is also an unforced
path (or cycle). By Lemma \ref{l4}, no unforced cycle is connected by an
unforced path to an unforced cycle. Therefore no cycle of unstable
gates is connected by a path of unstable gates to a cycle of unstable
gates. Then the size of any state cycle of $\langle\tilde{D},\tilde{b}\rangle$
is at most the least common multiple of the periods of all the cycles of
unstable gates. A cycle consisting of $s$ unstable gates has period $t$
or $2t$ for some factor $t$ of
$s$. Let $m$ be the size of the largest cycle of unstable gates. Then
the size of any state cycle of $\langle\tilde{D},\tilde{b}\rangle$ is
bounded by twice the least common multiple of all the natural numbers
less than or equal to $m$.

By the Prime Number Theorem (see \cite{t}), the number of primes less
than or equal to $m$ is asymptotic to $m\log e/\log m$. Then the least common
multiple of all the numbers less than or equal to $m$ is bounded by
$m^{1.45m/\log m}$, and
the size of any state cycle of $\langle \tilde{D},\tilde{b} \rangle$ is at most
$$
2m^{1.45m/\log m} \leq 2^{12\log\log m/\vep(n)}
$$
by Lemma \ref{l3}.
\hfill $\Box$
\begin{corollary}
Assume $\vep(n) \geq a\log\log n/\log n$ for some constant $a$. Then
there is a constant $c$ such that
\begin{eqnarray*}
\pr(\langle \tilde{D},\tilde{b}\rangle \mbox{\rm  \ has a tail longer than }
  c(\log n)^2/\log\log n) & = & o(1)\mbox{\rm , and} \\
\pr(\langle \tilde{D},\tilde{b}\rangle \mbox{\rm  \ has a state cycle larger
than }
n^c) & = & o(1).
\end{eqnarray*}
In particular, if $a > 12$, then we can take $c < 1$.
\end{corollary}
\section{Discussion} There are a number of problems suggested by our results.
An
immediate question is whether small upper bounds on tail length and state cycle
size can be proven for Kauffman's model. More generally, the effect of using
other distributions on the boolean functions should be investigated. Possibly
the uniform distribution, where each boolean function has probability $1/16$,
is
a threshold between chaotic and antichaotic behavior. That is, changing the
probability of a certain type of function from less than $1/16$ to greater than
$1/16$ may alter drastically the stability of the network. It may be meaningful
to
group the functions into several categories, such as constant functions,
functions depending on one argument, canalyzing functions depending on two
arguments, and non-canalyzing functions.

Consequences of using functions with more than two arguments should also be
studied.
Compared to the two argument functions, the three argument functions have a
much
smaller proportion of canalyzing functions, and there may be another threshold
involving the indegree of the gates. A network with a significant number of
three input gates is likely to have more non-canalyzing functions, and these
would be less likely to be stable.
\section{Acknowledgement} The author thanks Dr.\ Alan Woods of the
University of Western Australia for ideas and stimulating conversations
that improved this article.

\end{document}